\documentclass{article}
\usepackage{ijcai24}

\usepackage{times}
\usepackage{soul}
\usepackage{url}
\usepackage[hidelinks]{hyperref}
\usepackage[utf8]{inputenc}
\usepackage[small]{caption}
\usepackage{graphicx}
\usepackage{amsmath}
\usepackage{amsthm}
\usepackage{booktabs}
\usepackage{algorithm}
\usepackage{algorithmic}
\usepackage[switch]{lineno}
\usepackage{amsfonts}
\usepackage{color}
\usepackage{multirow}
\usepackage{xcolor,colortbl}
\usepackage{color, colortbl}
\definecolor{tabhighlight}{HTML}{e5e5e5}
\usepackage{diagbox}

\title{Vision-Informed Flow Image Super-Resolution with Quaternion Spatial Modeling and Dynamic Flow Convolution}

\author{
    Qinglong Cao\textsuperscript{\rm 1,2},
    Zhengqin Xu\textsuperscript{\rm 1},
    Chao Ma\textsuperscript{\rm 1},
    Xiaokang Yang\textsuperscript{\rm 1}
    Yuntian Chen\textsuperscript{\rm 2}\thanks{Corresponding Author},
   \\ $^1$ MoE Key Lab of Artificial Intelligence, AI Institute, Shanghai Jiao Tong University, China\\
$^2$ Ningbo Institute of Digital Twin, Eastern Institute of Technology, Ningbo, China\\
{\tt\small \{caoql2022, fate311, chaoma, xkyang\}@sjtu.edu.cn, ychen@eitech.edu.cn}
}

\begin{document}

\maketitle

\begin{abstract}
Flow image super-resolution (FISR) aims at recovering high-resolution turbulent velocity fields from low-resolution flow images. Existing FISR methods mainly process the flow images in natural image patterns, while the critical and distinct flow visual properties are rarely considered. This negligence would cause the significant domain gap between flow and natural images to severely hamper the accurate perception of flow turbulence, thereby undermining super-resolution performance. To tackle this dilemma, we comprehensively consider the flow visual properties, including the unique flow imaging principle and morphological information, and propose the first flow visual property-informed FISR algorithm. Particularly,  different from natural images that are constructed by independent RGB channels in the light field, flow images build on the orthogonal UVW velocities in the flow field. To empower the FISR network with an awareness of the flow imaging principle, we propose quaternion spatial modeling to model this orthogonal spatial relationship for improved FISR. Moreover, due to viscosity and surface tension characteristics, fluids often exhibit a droplet-like morphology in flow images. Inspired by this morphological property, we design the dynamic flow convolution to effectively mine the morphological information to enhance FISR. Extensive experiments on the newly acquired flow image datasets demonstrate the state-of-the-art performance of our method. Code and data will be made available. 
\end{abstract}

\section{Introduction}
Flow is a chaotic, spatio-temporal multi-scale nonlinear phenomenon that is ubiquitously found in our world. The precise measurement of flow holds significant implications across diverse domains, including but not limited to weather forecasting~\cite{bi2023accurate,zhang2023skilful}, energy assessment~\cite{juan2022cfd}, building design~\cite{zhong2022recent}, and hemodynamic analysis~\cite{arvidsson2022hemodynamic}. However, obtaining accurate measurements or simulations of flow with sufficiently high resolution typically requires substantial costs. Thus, flow image super-resolution has become a critical fluid dynamics task, which focuses on recovering high-resolution flow images from the counterpart low-resolution turbulent velocity fields.

\begin{figure}[t]
	\begin{center}
		\includegraphics[width=1.0\linewidth]{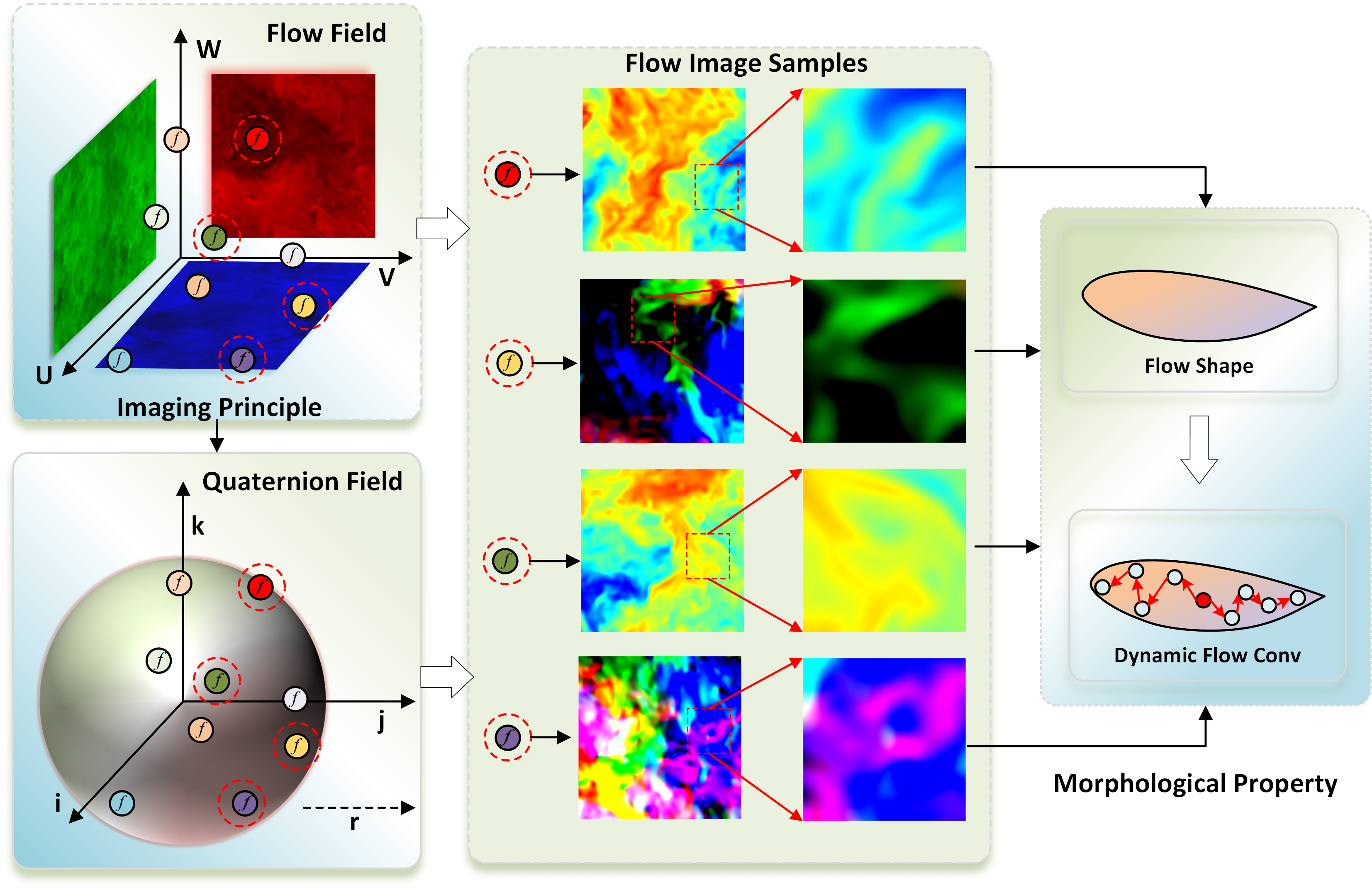}
	\end{center}
	\caption{Visual properties of flow images. Imaging Principle: the UVW velocities of the flow field that are set as RGB channels for imagery could be directly transferred into the orthogonal quaternion field. Morphological Property: The generalized droplet-like flow shape could be better analyzed by our designed dynamic flow convolution to provide the essential morphological information.  }
	\label{fig1}
\end{figure}

With the development of deep learning technology~\cite{simonyan2014very,he2016deep}, many deep learning-based approaches have been applied to handle the FISR problem~\cite{fukami2019super,liu2020deep,fukami2021machine}. For instance,  some researchers directly applied the ESRGAN~\cite{wang2018esrgan} to flow image super-resolution in a transfer learning manner~\cite{yu2022three}. Generative adversarial network (GAN)~\cite{goodfellow2020generative} has also been leveraged to achieve super-resolution reconstruction of turbulent flow fields at various Reynolds numbers~\cite{yousif2022super}. Moreover, through introducing physical constraints, Bao \textit{et al.}~\cite{bao2022physics} proposed a physics-guided neural network for reconstructing frequent flow images from sparse low-resolution data by enhancing its spatial resolution and temporal frequency. Following this pattern, physics-informed generative adversarial networks~\cite{li2022using} were proposed to perform super-resolution for multiphase fluid simulations. Harnessing the exceptional global perception ability of the transformer~\cite{vaswani2017attention}, Xu \textit{et al.}~\cite{xu2023super} proposed a transformer-based network to achieve FISR with better deep features. 

Despite advancements made by existing methods, they mainly process the flow images in the traditional natural image patterns, and the crucial flow visual properties including the distinct flow imaging principles and the flow morphological information are rarely considered. Independent RGB channels in the light field construct the natural images, yet the orthogonal UVW velocities in the flow field build the flow images. Despite its apparent randomness, turbulent flow maintains a consistent morphology governed by the Navier-Stokes (NS) equations, unlike the diverse shapes seen in natural objects. Consequently, the oversight of the visual properties would widen the domain gap and deteriorate the perception of flow turbulence, leading to a significant degradation in super-resolution performance. 


To address this challenge, we conduct a comprehensive analysis of the visual properties inherent in flow images and present the first flow visual-property-informed FISR algorithm, integrating quaternion spatial modeling and novel dynamic flow convolution. Specifically, as depicted in the left of Figure~\ref{fig1}, the UVW velocities of the flow field align with the \textbf{ijk} imaginary axes of the quaternion field. Recognizing this alignment, we advocate the incorporation of orthogonal spatial relations into the FISR network to enhance spatial modeling. Concretely, we map the UVW velocities into the quaternion field and leverage a quaternion network to mine the orthogonal spatial relations. Consequently, the UVW velocities of the flow field are individually assigned to the orthogonal imaginary axes of the quaternion field. This transformation facilitates a more effective exploration of inter- and intra-correlations through quaternion spatial modeling.

Meanwhile, since the distinctive viscosity and surface tension, could derived from the NS equation, flow turbulence often exhibits a droplet-like morphology, as illustrated in the right of Figure~\ref{fig1}. By incorporating this morphological information into the FISR network, the turbulence could be perceived more precisely and super-resolution performance could be boosted. Thus, motivated by this insightful observation, we proposed the dynamic flow convolution to adaptively capture the morphological information of fluids, where each grid position is decided by the previous grid position (Viscosity), and the offsets of positions are constrained by the flow shape (Surface Tension). Finally, through simultaneous quaternion spatial modeling and dynamic flow convolution, our proposed method successfully captures the visual properties and achieves superior super-resolution performance.

Our main contributions are summarized as follows:

\begin{itemize}
	\item To the best of our knowledge, this is the first work to introduce the visual properties of flows into the task of flow image super-resolution. It successfully narrows the domain gap and facilitates more precise perception of the chaotic turbulence.
 

	\item The proposed quaternion spatial modeling competently models the latent orthogonal relation within UVW velocities, and the inter- and intra-correlations in the transferred quaternion field are effectively explored.

    \item The proposed dynamic flow convolution successfully extracts the morphological information of fluids, which empowers the network with the knowledge of turbulence appearance.

	\item We extensively validate our method on the flow image datasets and conduct ablation studies to examine its characteristics. Experimental results show that our method achieves state-of-the-art performance.
	
\end{itemize}

\section{Related Work}
\textbf{Image Super-Resolution.} Image super-resolution is a fundamental and critical problem in computer vision. Plenty of efforts have been devoted to this area and many researchers have achieved significant improvements. For instance, RCAN~\cite{zhang2018image} incorporated the deep residual channel attention to push the network focus on the interdependencies among channels. Moreover, SwinIR~\cite{liang2021swinir} first introduced the famous Transformer architecture into the image super-resolution network and successfully boosted the performance. Following this pattern,  TTSR~\cite{yang2020learning} proposed a texture-focused transformer network to perform image super-resolution with a hard-attention module for texture transfer and a soft attention module for texture synthesis. However, these methods still utilize the transformer in a single dimension, thus DAT~\cite{chen2023dual} aggregates features across spatial and channel dimensions, and performed the image super-resolution in the inter-block and intra-block dual manner. Attempting to solve the uncertainty challenge in Image Super-resolution, DDL~\cite{liu2023spectral} combined with Bayesian approaches estimated spectral uncertainty accurately. Assuming the natural images as the long-tailed pixel distribution, Gou \textit{et al.}~\cite{gou2023rethinking} introduced a static and
a learnable structure prior to re-balance the gradients from the pixels in the low- and high-frequency region. Focusing on improving the efficiency of image super-resolution, FSR~\cite{li2023fsr} accelerated super-resolution networks by considering data characteristics in the frequency domain. HPUN~\cite{sun2023hybrid} leveraged pixel-unshuffled downsampling and self-residual depthwise separable convolutions to construct a lightweight image super-resolution network. Recently, aiming at further exploiting the super-resolution potential of transformers, HAT~\cite{chen2023activating} leveraged both channel attention and window-based self-attention schemes to activate more pixels with a better super-resolution performance.

\textbf{Flow Image Super-Resolution.} Flow image super-resolution focuses on recovering high-resolution turbulent flows from grossly coarse flow images. Existing researches tend to transfer the natural vision-based methods to process the flow images. Particularly, a static convolutional neural network (SCNN), and a multiple temporal paths convolutional neural network are simultaneously leveraged in ~\cite{liu2020deep} to capture spatial and temporal information for better super-resolution performance. Apart from the efficient convolutional network, generative adversarial networks(GAN) are also adopted to perform the flow image super-resolution. For example, these researches~\cite{yousif2022super,yousif2022super,deng2019super} respectively utilize the GAN-based networks to perform effective super-resolution. Focusing on utilizing the physical knowledge to boos the super-resolution~\cite{bode2019using}, this work~\cite{bao2022physics} designed a partial differential equation (PDE)-based recurrent unit for capturing underlying temporal processes and incorporated additional physical constraints to supervise the learning. Since the scarcity and huge cost of labeled data, unsupervised learning ~\cite{kim2021unsupervised} has also been attempted to perform the flow image super-resolution. However, these methods all directly transfer the natural vision-based methods to handle the flow image super-resolution, the domain gap between flow and natural would seriously hamper the accurate perception of flow turbulence and further demote the super-resolution performance. To tackle this problem, we proposed the first vision-informed flow image super-resolution algorithm.

\section{Preliminary}
\textbf{Quaternion Networks.}
We first introduce the quaternion background knowledge for the following quaternion spatial modeling. In four-dimensional space, a quaternion $Q$ extends a hyper-complex number and can be expressed as follows:
\begin{equation}
 Q= r1+ x\textbf{i} + y\textbf{j} + z\textbf{k},
\end{equation}
where r, x, y, and z are real numbers, and 1, $\textbf{i},\textbf{j}$, and $\textbf{k}$ are the quaternion unit basis. The real part of $Q$ is denoted by r, while $x\textbf{i} + y\textbf{j} + z\textbf{k}$ is the imaginary or vector part. The Hamilton product $\otimes$ of two quaternions $Q_1$ and $Q_2$ is computed as:
\begin{equation}
\begin{split}
{Q_1} \otimes {Q_2} = & ({r_1}{r_2} - {x_1}{x_2} - {y_1}{y_2} - {z_1}{z_2}) \\ & + ({r_1}{x_2} + 
 {x_1}{r_2} + {y_1}{z_2} - {z_1}{y_2})\textbf{i} \\ & + ({r_1}{y_2} - {x_1}{z_2} +
 {y_1}{r_2} + {z_1}{x_2})\textbf{j} \\ & + ({r_1}{z_2} + {x_1}{y_2} - {y_1}{x_2} + {z_1}{r_2})\textbf{k} ,
\end{split}
\end{equation}
The split activation function $\alpha$  works on quaternion is defined as:
\begin{equation}
\alpha (Q) = f(r)1 + f(x)\textbf{i} + f(y)\textbf{j} + f(z)\textbf{k},
\end{equation}
with $f$ corresponding to any standard and real-valued activation function.

Let $\gamma^{l}_{ab}$ and $S^{l}_{ab}$ respectively be the quaternion output and the pre-activation quaternion output at layer $l$ and at the indexes $(a,b)$ of the new feature map, and $\omega$ be the quaternion-valued weight filter map of size $K \times K$. A formal definition of the convolution process is defined as:

\begin{equation}
S_{ab}^l = \sum\limits_{c = 0}^{K - 1} {\sum\limits_{d = 0}^{K - 1} {{w^l} \otimes \gamma _{(a + c)(b + d)}^{l - 1}} } ,
\end{equation}
\begin{equation}
\gamma _{ab}^l = \alpha (S_{ab}^l) ,
\end{equation}
where $a$ can be any split activation function. In this extent, a traditional 2D quaternion convolutional layer, with a kernel that contains $f$ feature maps, is split into four parts: the first part is equal to r, the second one to $x\textbf{i}$, the third one to $y\textbf{j}$ and the last one to $z\textbf{k}$ of a quaternion $Q$.

\textbf{Deformable Convolution.} Preparing knowledge for the newly proposed dynamic flow convolution in this section.  A traditional convolution could be divided into two steps: 1) sampling using a regular grid $R$ over the input feature map $x$; 2) summation of sampled values weighted by $w$. The grid $R$ defines the receptive field size and dilation. Particularly, given the grid $R$:
\begin{equation}
R = \left\{ {\left( { - 1, - 1} \right),( - 1,0),...,(0,1),(1,1)} \right\},
\end{equation}
defines a $3 \times 3$ kernel with dilation 1.
For each location $p_0$ on the output feature map $y$, the convolution is computed as follow:
\begin{equation}
y({p_0}) = \sum\limits_{{p_n} \in R} {w({p_n})x\left( {{p_0} + {p_n}} \right)}, 
\end{equation}
In deformable convolution~\cite{dai2017deformable}, the regular grid $R$ is augmented with learnable offsets $\left\{ {\Delta {p_n}|n = 1,...,N} \right\}$, where $N = |R|$. Consequently, the deformable convolution is computed as follow:
\begin{equation}
y({p_0}) = \sum\limits_{{p_n} \in R} {w({p_n})x\left( {{p_0} + {p_n} + \Delta {p_n}} \right)} ,
\end{equation}
Clearly, the sampling is done at irregular offset locations ${p_n} + \Delta {p_n}$. However, since the offset $\Delta P_n$ is typically fractional, bilinear interpolation is often implemented:
\begin{equation}
\label{inter}
x(p) = \sum\limits_q {G(q,p) \cdot x(q)}, 
\end{equation}
where $p$ denotes an arbitrary location, $q$ enumerates all integral spatial locations in the feature map x, and $G( \cdot , \cdot )$ is the bilinear interpolation kernel. $G$ is a two-dimensional function, thus it is separated into two one-dimensional kernels:
\begin{equation}
G(q,p) = g({q_x},{p_x}) \cdot g({q_y},{p_y}),
\end{equation}
where $g(a,b) = \max (0,1 - |a - b|)$.
\section{Proposed Method}
\subsection{Method Overview}
\begin{figure*}[t]
	\begin{center}
		\includegraphics[width=1.0\linewidth]{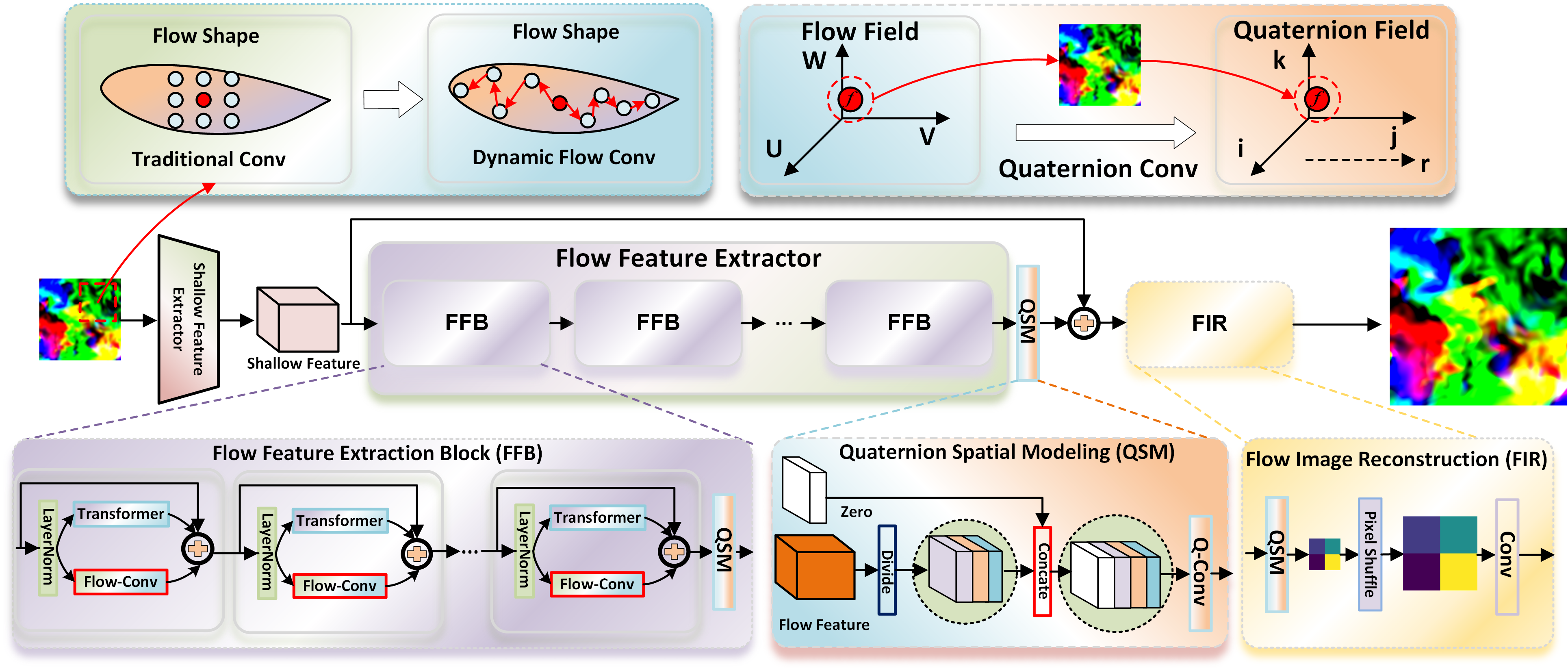}
	\end{center}
	\caption{The pipeline of our proposed vision-informed flow image super-resolution algorithm. The low-resolution flow images are first propagated into one convolution layer to generate the shallow feature. Then, through the sequential Flow Feature extraction Blocks (FFB),  the vision-informed flow features are acquired. Subsequently, through the Quaternion Spatial Modeling (QSM) and a deep residual layer, the flow features are inputted into the Flow Image Reconstruction (FIR) module to obtain the final high-resolution flow images.    }
	\label{fig2}
 \vspace{-2mm}
\end{figure*}

Inspired by the flow visual properties, we proposed the first vision-informed flow image super-resolution algorithm with quaternion spatial modeling and dynamic flow convolution. The overall architecture is shown in Figure~\ref{fig2}. More specifically, one convolution layer is first introduced to extract the shallow feature of low-resolution flow images. Subsequently, the shallow features are propagated into the flow feature extractor to extract the vision-informed flow feature. Particularly, in each flow feature extraction block, the swin-transformer layer~\cite{liu2021swin} is leveraged to capture the global information, the designed dynamic flow convolution is utilized to capture the local morphological information, and the quaternion convolution is leveraged at last layer to mine spatial orthogonal relation. This feature process lasts several times in each block (FFB), and multiple blocks consist of the flow feature extractor. After that, along with the quaternion spatial modeling (QSM), a deep residual layer is leveraged to avoid forgetting the shallow features. Finally, utilizing the quaternion convolution to model the spatial relations and the pixel shuffle layer to upsample the image, the network successfully outputs the high-resolution flow images. 


\subsection{Quaternion Spatial Modeling}
The UVW velocities of the flow field are respectively leveraged to set at RGB channels for imagery:
\begin{equation}
(U,V,W) \to (R,G,B),
\end{equation}
Thus, there exist spatial orthogonal relations among the set RGB channels. Yet, their orthogonal relations are rarely considered in previous FISR methods. Inspired by the quaternion networks~\cite{parcollet2018quaternion}, known for their effective orthogonal relation modeling and
powerful exploration of inter- and intra-correlations within
the quaternion hidden space, we perform quaternion spatial modeling on the UVW-based flow features. Given input features $F$, the features would be evenly divided along the channel dimension into three parts $F \in {\mathbb{R}^{H \times W \times C_f}} \to [{F_1},{F_2},{F_3}]$, where the first part $F_1 \in {\mathbb{R}^{H \times W \times \frac{{{C_f}}}{3}}}$ belongs to the $\textbf{i}$ axis, the second part $F_2 \in {\mathbb{R}^{H \times W \times \frac{{{C_f}}}{3}}}$ belongs to the  $\textbf{j}$ axis, and the third part $F_3 \in {\mathbb{R}^{H \times W \times \frac{{{C_f}}}{3}}}$ belongs to the $\textbf{k}$ axis:
\begin{equation}
{Q_f} = 0 + {F_1}\textbf{i} + {F_2}\textbf{j} + {F_3}\textbf{k},
\end{equation}
where $Q_f$ denotes the quaternion field for flow. Thus, to support the quaternion spatial modeling, we first construct a zero tensor $Z_o$ with the same size as $F_1$. Subsequently, given the quaternion convolution layer as $Q_t$, the process of quaternion spatial modeling would be computed as:
\begin{equation}
{F_q} = {Q_t}([{Z_0},{F_1},{F_2},{F_3}]),
\end{equation}
where $[ \cdot, \cdot ]$ denotes the concatenation operation, and $F_q$ is the output feature of the quaternion spatial modeling (QSM). 

\subsection{Dynamic Flow Convolution}
To mine the morphological information of fluid, we creatively proposed the dynamic flow convolution. Formally, given the standard 2D convolution coordinates as $K$, the central coordinate is ${K_i} = ({x_i},{y_i})$, which corresponds to the red grid in the dynamic flow conv of Figure~\ref{fig2}. A $3 \times 3$ kernel $K$ with dilation 1 is expressed as:
\begin{equation}
K = \{ (x - 1,y - 1),(x - 1,y), \cdot  \cdot  \cdot ,(x + 1,y + 1)\} ,
\end{equation}

In dynamic flow convolution (DFC), the standard convolution kernel is decoupled in the direction of the x-axis and the y-axis. Taking the convolution kernel of size 9 as an example, the specific position of each grid in $K$ is represented as ${K_{i \pm t}} = ({x_{i \pm t}},{y_{j \pm t}})$, where 
$t = \left\{ {0,1,2,3,4} \right\}$ denote the distance from the center grid. Considering the viscosity, the position offset of each grid should be decided by the position offsets of all previous grids, namely a chain decision procedure. Particularly, in the direction of the x-axis, the shift of position is computed as:
\begin{equation}
\label{ekx}
{K_{i \pm t}} = \left\{ {\begin{array}{*{20}{c}}
{({x_{i + t}},{y_{i + t}}) = ({x_i} + t,{y_i} + \sum\nolimits_i^{i + t} {\Delta y} )}\\
{({x_{i - t}},{y_{i - t}}) = ({x_i} - t,{y_i} + \sum\nolimits_{i - t}^i {\Delta y} )}
\end{array}} \right. ,
\end{equation}
Correspondingly, in the direction of the y-axis, the shift of position is computed as:
\begin{equation}
\label{eky}
{K_{j \pm t}} = \left\{ {\begin{array}{*{20}{c}}
{({x_{j + t}},{y_{j + t}}) = ({x_j} + \sum\nolimits_j^{j + t} {\Delta x} ,{y_j} + t)}\\
{({x_{j - t}},{y_{j - t}}) = ({x_j} + \sum\nolimits_{j - t}^j {\Delta x} ,{y_j} - t)}
\end{array}} \right. 
\end{equation}

Simultaneously, considering the surface tension, there is always a wider side (left or right) in the flow turbulence, thus the learned offsets should be appropriately constrained. More specifically, taking the left as an example, the offset of the center grid remains the same, while the true flow offsets $[\Delta_f x, \Delta_f y]$ of other grids  would be decided by:

\begin{equation}
\begin{aligned}
{\Delta _f}{x_{i + 1}} &= sgn(\Delta {x_{i + 1}})Max(|\Delta {x_{i + 1}}|,|\Delta {x_i}|)\\
{\Delta _f}{x_{i - 1}} &= sgn(\Delta {x_{i - 1}})Min(|\Delta {x_{i - 1}}|,|\Delta {x_i}|),
\end{aligned}
\end{equation}

\begin{equation}
\begin{aligned}
{\Delta _f}{y_{j + 1}} &= sgn(\Delta {y_{j + 1}})Max(|\Delta {y_{j + 1}}|,|\Delta {y_j}|)\\
{\Delta _f}{y_{j - 1}} &= sgn(\Delta {y_{j - 1}})Min(|\Delta {y_{j - 1}}|,|\Delta {y_j}|),
\end{aligned}
\end{equation}

where $sgn(\cdot)$ is the sign function, which gives the sign of input. Thus, in order to implement the constraints, $\Delta y$ in Eq.~\ref{ekx} and $\Delta x$ in Eq.~\ref{eky} should be respectively updated as $\Delta_f x$ and $\Delta_f y$. Since the offset normally would be fractional, we also adopt the bilinear interpolation in Eq.~\ref{inter}. If the right side is wider, the constraints should be correspondingly changed.

Since the two patterns of the wider side tend to simultaneously exist in the random turbulence, both patterns of constraints are leveraged in our method to obtain two kinds of flow perception features $\{F_L, F_R\}$, where $F_L$ is obtained through the left-pattern dynamic flow convolution, and $F_R$ is obtained by the right-pattern dynamic flow convolution. Afterwards, both kinds of flow perception features are fused to obtain the final flow perception features. Suppose the input feature is $F$, the computation process is given as:
\begin{equation}
{F_L} = DF{C_{left}}(F), {F_R} = DF{C_{right}}(F) ,
\end{equation}
\begin{equation}
{F_p} = Conv[{F_L},{F_R}] ,
\end{equation}
where $F_p$ denotes the final flow perception feature, and $[\cdot,\cdot]$ means the concatenation operation.

\subsection{Flow Feature Extractor}
Given the low-resolution flow image ${I_{LR}} \in {\mathbb{R}^{H \times W \times {\rm{3}}}}$, we first leverage one convolutional layer to extract the shallow features ${F_{\rm{0}}} \in {\mathbb{R}^{H \times W \times C}}$, where $C$ denotes the channel number. Then the shallow features are propagated into the flow feature extractor. A series of sequential flow feature extractor blocks (FFB) consists of the flow feature extractor. More specifically, each feature extractor block also contains a series of feature extraction units. In each unit,  the features are first inputted into the  LayerNorm (LN) layer, and then the standard swin-transformer (SWINT) layer is leveraged to extract the global information. Simultaneously, the proposed dynamic flow convolution (DFC) is utilized to capture the critical morphological information. Then a residual connection is also adopted to avoid forgetting. Formally, given the input feature $X$, the process of the feature extraction unit ($U_f$) is computed as follows:
\begin{equation}
{X_N}{\rm{ = }}LN(X) ,
\end{equation}
\begin{equation}
{X_{out}} = SWINT({X_N}) + DFC({X_N}) + X ,
\end{equation}
where $X_{out}$ is the output of every unit. Note that the quaternion spatial modeling (QSM) is added at the last layer of each FFB to enhance the perception of spatial relation during the feature extraction process. Given the input of the FFB as $Z$, the computation of FFB could be formally given as:
\begin{equation}
{Z_{out}} = QSM({U_f}({U_f}(...{U_f}(Z)...))) ,
\end{equation}
where $Z_{out}$ is the output of each FFB. Correspondingly, the whole computation process of flow feature extractor could be given as:
\begin{equation}
{F_f} = FFB(FFB(...FFB({F_0})...)) ,
\end{equation}
where $F_f$ denotes the extracted flow features with sufficient flow global perception information and local morphological information. 

\subsection{Flow Image Reconstruction}
Before going into the flow image reconstruction module, the extracted flow features first go through a quaternion spatial modeling layer and a deep residual layer from the shallow features to prevent forgetting the low-level knowledge:
\begin{equation}
\widehat F_f = QSM({F_f}) + {F_0} ,
\end{equation}
Subsequently, to empower the reconstruction process with the flow spatial orthogonal relations, a quaternion spatial modeling layer is added, and a Pixel-Shuffle (PS) layer is utilized to upscale the image. Finally, a convolution layer is utilized to reconstruct the predicted high-resolution image $\widehat I_{HR}$:
\begin{equation}
{\widehat I_{HR}} = Conv(PS(QSM({\widehat F_f}))) ,
\end{equation}
We directly use $L_1$ loss to optimize the network parameters.

\begin{table*}[t!]
	\centering
	\caption{  Quantitative comparison with state-of-the-art methods on flow images datasets.  The best results are shown in bold. $\uparrow$ and $\downarrow$ denote larger and smaller is better.}
	\scalebox{1.0}{
		\begin{tabular}{cc|cccc|cccc}
			\hline
				\multicolumn{2}{c|}{\multirow{2}{*}{Method}}&  \multicolumn{4}{c|}{Single-Velocity}&\multicolumn{4}{c}{Multi-Velocities}  \\
			\multicolumn{2}{c|}{}     
			& PSNR$\uparrow$  & SSIM$\uparrow$  & RMSE$\downarrow$  & MAE$\downarrow$  & PSNR$\uparrow$ & SSIM$\uparrow$ & RMSE$\downarrow$  & MAE$\downarrow$  \\ 
			\hline \hline
			\multicolumn{10}{c}{$\times$2}  \\ \hline  
			\multicolumn{2}{c|}{RCAN~\cite{zhang2018image}}   & 40.32 & 0.9801 & 1.9901 & 0.9430 & 40.85 &0.9847 & 2.1900  & 0.5617 \\ 
			\multicolumn{2}{c|}{SwinIR~\cite{liang2021swinir}}   & 42.51 & 0.9836 & 1.9976 & 0.9880 & 35.78 &0.9872 & 4.2083 & 1.6011  \\ 
			\multicolumn{2}{c|}{HAT~\cite{chen2023activating}}    & 42.33 & 0.9880 & 1.9665 & 0.9389 & 43.40 &0.9905 & 1.5685 & 0.5557 \\ 
			\multicolumn{2}{c|}{DAT~\cite{chen2023dual}} &42.61  &0.9887 &1.9146 &0.8566 &43.22 &0.9938   &1.7829 &0.6369  \\ 
   	 \rowcolor{tabhighlight} \multicolumn{2}{c|}{Ours} &\textbf{43.37} &\textbf{0.9904}  &\textbf{1.7769}  &\textbf{0.8456} &\textbf{45.81} &\textbf{0.9966} &\textbf{1.2527}  &\textbf{0.4408} \\ 
			\hline
			\hline
			\multicolumn{10}{c}{$\times$4}  \\ \hline
			\multicolumn{2}{c|}{RCAN~\cite{zhang2018image}}   & 38.12  & 0.9760 & 4.6180 & 2.5321 & 38.77 &0.9821 & 3.0114  & 1.1268 \\ 
			\multicolumn{2}{c|}{SwinIR~\cite{liang2021swinir}}   & 33.90 & 0.9711 & 5.1942 & 2.8949 & 26.34 &0.9350 &12.4827 & 4.9372  \\ 
			\multicolumn{2}{c|}{HAT~\cite{chen2023activating}}    & 38.61 & 0.9798 & 3.8820 & 1.8589 & 37.60 &0.9877 & 3.4148 & 0.9061 \\ 
			\multicolumn{2}{c|}{DAT~\cite{chen2023dual}} &41.94  &0.9835 &2.8071 &1.4980 &39.47 &0.9867   & 2.7506  & 0.7331  \\ 
      \rowcolor{tabhighlight} \multicolumn{2}{c|}{Ours} &\textbf{42.93}  &\textbf{0.9896} &\textbf{1.8223}  &\textbf{0.9689} &\textbf{44.82} &\textbf{0.9921} &\textbf{1.3201} &\textbf{0.4814} \\
   \hline \hline
			\multicolumn{10}{c}{$\times$8}  \\ \hline
			\multicolumn{2}{c|}{RCAN~\cite{zhang2018image}}    & 37.60 & 0.9650 & 5.6141 & 4.8593 & 34.16 & 0.9722 & 5.4032 & 1.3921  \\  
			\multicolumn{2}{c|}{SwinIR~\cite{liang2021swinir}}   & 27.12 & 0.9180  & 11.3352 & 6.5020 & 21.40 &0.8578 & 22.0178 & 9.4719  \\ 
   			\multicolumn{2}{c|}{HAT~\cite{chen2023activating}} &38.19 &0.9628 &5.3012 &2.7386 &28.76 &0.9547 &9.5054 &3.5754 \\ 
            \multicolumn{2}{c|}{DAT~\cite{chen2023dual}} &41.04 &0.9781 &3.1715 &1.7115 &35.36 &0.9855 &4.4284 &1.6365 \\ 
     \rowcolor{tabhighlight}  \multicolumn{2}{c|}{Ours} &\textbf{42.34} &\textbf{0.9863} &\textbf{1.9562} &\textbf{0.9981} &\textbf{37.56} &\textbf{0.9876} &\textbf{3.1557} &\textbf{0.8506} \\
   \hline 
	\end{tabular}}
	
	\label{tab1}	
  \vspace{-3mm}
\end{table*}

\begin{figure*}[t]
	\begin{center}
		\includegraphics[width=0.9\linewidth]{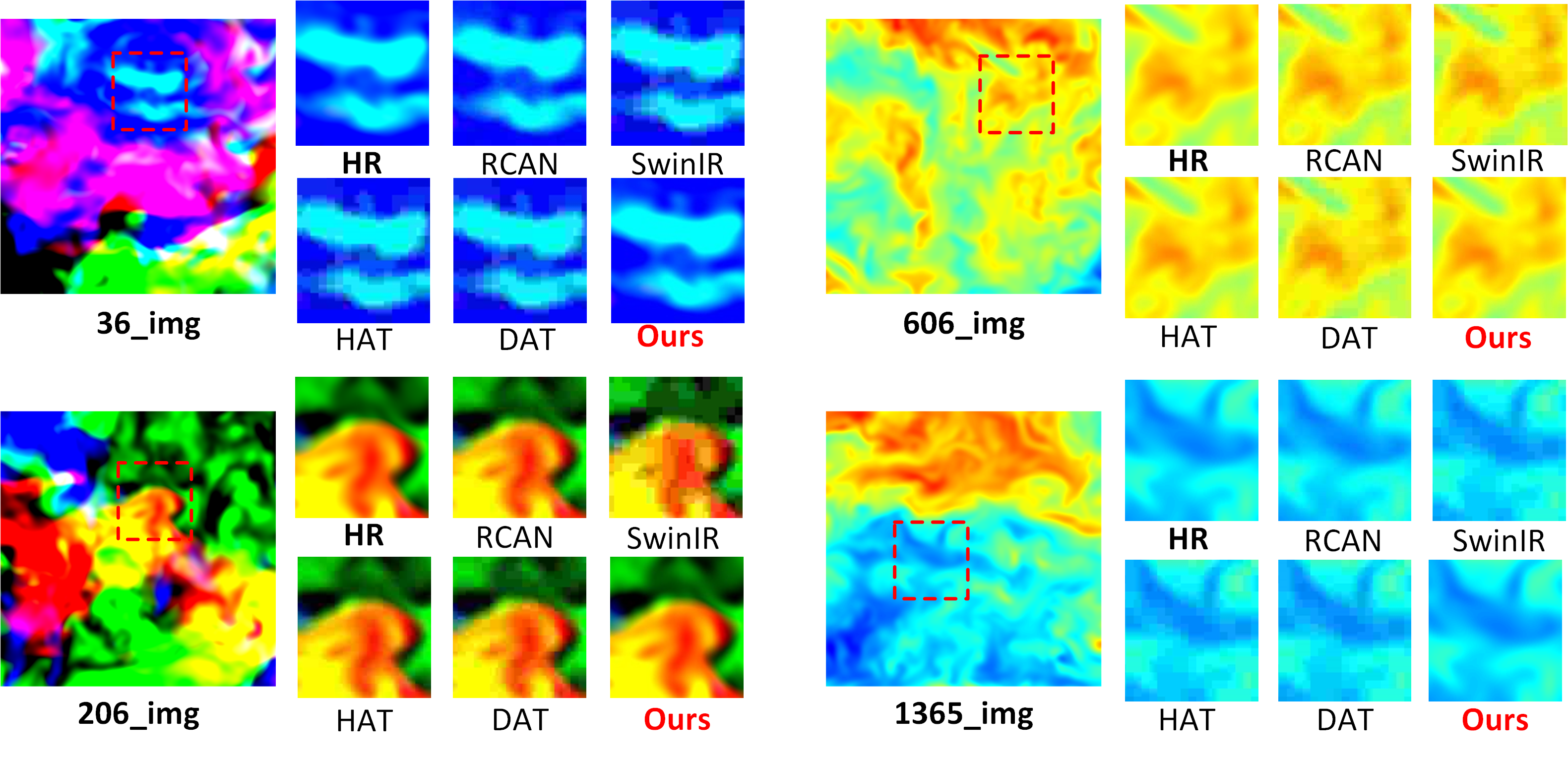}
	\end{center}
  \vspace{-2mm}
	\caption{The visual comparisons with advanced methods.  HR: the high-resolution flow images. Ours: the results of our proposed method.  }
	\label{fig3}

\end{figure*}

\section{Experiments}
\subsection{Experimental Settings}
\textbf{Data and Evaluation.} The utilized flow images dataset comes from the direct numerical simulation (DNS) data with the commonly used Navier-Stokes (NS) equations. Particularly, the flow images are acquired from simulating incompressible isotropic turbulence undergoing natural decay at a Taylor Reynolds number of approximately $Re_{\lambda} \approx 250$. We will release the dataset for further research. The high-resolution flow images are provided with a resolution of $512 \times 512$ and the low-resolution flow images with resolutions of $256 \times 256$, $128 \times 128$, and $64 \times 64$ are offered to respectively implement the $\times 2$,  $\times 4$, $\times 8$ super-resolution tasks. Since the UVW velocities (RGB channels) of flow turbulence could be respectively or simultaneously analyzed. Thus, two settings including single-velocity and multi-velocities are adopted for the research. For the single-velocity setting, we have $29491$ images for training and $7372$ images for testing. For the multi-velocities setting, $9750$ training images and $2475$ testing images are provided.  Following previous research, we adopt the PSNR and SSIM~\cite{wang2004image} as the evaluation metrics. Meanwhile, since the flow images are the numerical solutions for precise application, the RMSE and MAE metrics are also adopted to testify the performance. To better analyze the numerical error, RMSE and MAE are both $\times 255$ for comparison.

\textbf{Implementation Details.} Two NVIDIA A800 GPUs with the pytorch framework are leveraged to conduct the experiments. We utilize the Adam with a learning rate of 3e-4 and batch size of 12 to optimize the parameters. The training process lasts for 80000 iterations. The ema decay is set as 0.999, and the low-resolution images are set as 64 for saving computation space. The layers of feature extraction unit and flow feature extraction block are both set as 6.

\subsection{Comparison with State-of-the-art Methods}
\textbf{Quantitative Results.} Table~\ref{tab1} showcases a comprehensive performance comparison between our method and other state-of-the-art super-resolution techniques. The visualization results are based on the $\time 4$ task. Our approach consistently outperforms all previous methods, demonstrating clear superiority across both single-velocity and multi-velocities settings across all metrics. Notably, the highest PSNR performance is achieved at the $\times 2$ task within the multi-velocities setting, reaching 45.81, marking a substantial 2.41 dB improvement. Additionally, the most significant PSNR boost of 5.35 dB occurs at the $\times 4$ task within the multi-velocities setting. Specifically, at the $\times 2$ task, the most remarkable improvement is observed in the MAE metric under the multi-velocities setting, reducing the MAE from 0.5557 to 0.4408. For the $\times 4$ task, the most impressive performance gain is reflected in the RMSE, again within the multi-velocities setting, resulting in a nearly 1.5 RMSE reduction. Furthermore, at the $\times 8$ task, the most significant performance improvement is again seen within the multi-velocities setting, leading to a 1.2727 RMSE drop. It's worth noting an intriguing observation that the performance decrease from $\times 2$ task to $\times 4$ task appears to be smaller compared to the decrement from $\times 2$ task to $\times 8$ task. This suggests that the challenges in flow super-resolution do not follow a linear progression but exhibit a more rapid increase. Additionally, the substantial improvements observed in the multi-velocities setting highlight the importance of considering spatial relations and morphological information in simultaneously analyzing UVW velocities.

\textbf{Qualitative Results.} To comprehensively assess the effectiveness of our proposed method, we present visual comparisons between our approach and other state-of-the-art methods in Figure~\ref{fig3}. Evidently, our method excels in accurately capturing turbulence appearances, producing super-resolution results with finer visual details. Notably, even in situations where turbulence shapes vary, and flow directions are intricate, our method consistently maintains well-defined turbulence shapes and more precisely constructs adhesion areas. We attribute these superior results to the successful extraction of orthogonal spatial relations and the effective representation of turbulence appearances in our method. Upon closer examination of the visualization results, it becomes apparent that SwinIR exhibits notably poor performance, even worse than the results obtained using the purely convolutional neural network-based RCAN. This aligns with the quantitative findings presented in Table~\ref{tab1}. This observation suggests that while transformers excel at capturing global flow information, local information proves to be more critical for flow image super-resolution. Pure transformers may struggle to appropriately handle the intricacies of flow image super-resolution. This insight could serve as a valuable inspiration for future research, encouraging a focused exploration of methods to address this specific challenge.

\begin{table}[t]
	\centering
	\caption{ Ablation study on quaternion spatial modeling (QSM) and dynamic flow convolution (DFC).}
	\scalebox{1.0}{
		\begin{tabular}{c|cccc}
			\hline
			\multicolumn{1}{c|}{Methods}  &\multicolumn{1}{c}{PSNR$\uparrow$}  &\multicolumn{1}{c}{SSIM$\uparrow$}  &\multicolumn{1}{c}{RMSE$\downarrow$} &\multicolumn{1}{c}{MAE$\downarrow$}\\
			\hline
      	\multicolumn{1}{c|}{Baseline} &42.33	&0.9880	&1.9665 &0.9389 \\
   		\multicolumn{1}{c|}{Baseline+QSM} &43.21	&0.9889	&1.7917 &0.8619\\
            \multicolumn{1}{c|}{Baseline+DFC} &42.98	&0.9897 	&1.8328 &0.8676\\
   		\multicolumn{1}{c|}{Ours}   &\textbf{43.37}	&\textbf{0.9904}	&\textbf{1.7769} &\textbf{0.8456}\\
			\hline	
	\end{tabular} }
	\label{table2}
  \vspace{-2mm}
\end{table}

\begin{table}[t]
	\centering
	\caption{ Ablation study on the design of dynamic flow convolution (DFC). NDC: Normal Deformable Convolution, ADFC: Adaptive-DFC, LDFC: Left-DFC, RDF: Right-DFC. }
	\scalebox{1.0}{
		\begin{tabular}{c|cccc}
			\hline
			\multicolumn{1}{c|}{Methods}  &\multicolumn{1}{c}{PSNR$\uparrow$}  &\multicolumn{1}{c}{SSIM$\uparrow$}  &\multicolumn{1}{c}{RMSE$\downarrow$} &\multicolumn{1}{c}{MAE$\downarrow$}\\
			\hline
      	\multicolumn{1}{c|}{Baseline} &42.33	&0.9880	&1.9665 &0.9389 \\
        \multicolumn{1}{c|}{Baseline+NDC} &42.78	&0.9888 	&1.8718 &0.8714\\
   	\multicolumn{1}{c|}{Baseline+LDFC} &43.15	&0.9882	&1.8617 &0.8637\\
        \multicolumn{1}{c|}{Baseline+RDFC} &43.17	&0.9891 	&1.8409 &0.8619\\
        \multicolumn{1}{c|}{Baseline+ADFC} &43.11 	&0.9881 	&1.8434  &0.8572\\
   		\multicolumn{1}{c|}{Ours DFC}   &\textbf{43.21}	&\textbf{0.9897}	&\textbf{1.8328} &\textbf{0.8566}\\
			\hline	
	\end{tabular} }
	\label{table3}
\end{table}

\subsection{Ablation Study}
We set a series of ablation experiments to demonstrate the effectiveness of our proposed model. The experiments are conducted in the $\times 2$ task under the single-velocity setting.

\textbf{Visual Properties.} Our proposed method leverages quaternion spatial modeling to extract spatial orthogonal relations and utilizes dynamic flow convolution to provide crucial morphological information. To showcase the effectiveness of these two modules, we conducted experiments, and the results are presented in Table~\ref{table2}. It is evident from the table that both quaternion spatial modeling and dynamic flow convolution contribute to performance gains. Notably, the incorporation of dynamic flow convolution yields more substantial performance improvements compared to quaternion spatial modeling. This observation highlights that morphological information plays a more significant role in enhancing flow image super-resolution than the spatial orthogonal relations mined by quaternion spatial modeling.

\textbf{Design of Dynamic Flow Convolution.} We leverage the dynamic flow convolution to empower the network with the knowledge of turbulence appearance. We compared our dynamic flow convolution (DFC) with normal deformable convolution, the left-pattern DFC, the right-pattern DFC, and the adaptive DFC.  The adaptive DFC refers to the direction oriented towards the higher learnable offset's direction. The experimental results, detailed in Table~\ref{table3}, demonstrate that various convolutional approaches contribute to performance improvements. However, our method consistently outperforms all alternatives, underscoring the efficacy of our proposed approach in capturing essential morphological information for enhanced performance in flow image super-resolution.

\begin{table}[t]
	\centering
	\caption{ Ablation study on the layer numbers of feature extraction unit (FEU) and flow feature extractor block (FFB). Utilizing the PSNR as the metric.}
	\scalebox{1.0}{
		\begin{tabular}{c|cccc}
			\hline
			\multicolumn{1}{c|}{\diagbox{FEU}{FFB}}  &\multicolumn{1}{c}{4}  &\multicolumn{1}{c}{5}  &\multicolumn{1}{c}{6} &\multicolumn{1}{c}{7}  \\
			\hline
   		\multicolumn{1}{c|}{4}  &43.18 &43.27	&43.32 &43.35 \\
      \multicolumn{1}{c|}{5} &43.28	&43.29	&43.36	&43.34	  \\
        \multicolumn{1}{c|}{6} &43.24	&43.32	&\textbf{43.37}	&43.33	  \\
        \multicolumn{1}{c|}{7}   &43.23	&43.26	&43.33	&43.34	\\
			\hline	
	\end{tabular} }
	\label{table4}
\end{table}

\textbf{Layer Numbers.} Since a series of blocks are adopted in the flow feature extractor to capture the flow features,  we conducted experiments to determine the optimal layer numbers. The results of these experiments are presented in Table~\ref{table4}. Notably, the table illustrates that the network attains its peak performance when both the layer numbers of FEU and FFB are set to 6. Thus the layer numbers are appropriately configured.

\section{Conclusion}
Existing flow image super-resolution methods primarily process flow images in natural image patterns. However, the critical flow visual properties, including imaging principles and morphological properties, are seldom considered and explored. This lack of consideration can significantly degrade super-resolution performance. To handle this problem, we proposed the vision-informed flow images super-resolution method with quaternion spatial modeling and dynamic flow convolution. Concretely, utilizing the quaternion spatial modeling, the orthogonal spatial relations between UVW velocities are well-mined with improved performance. The proposed dynamic flow convolution effectively captures the turbulence appearance and empowers the network with morphological information. We conduct extensive experiments on the flow image datasets, and compelling experimental results demonstrate the superiority of our proposed method.






\bibliographystyle{named}
\bibliography{ijcai24}

\end{document}